\documentclass[aip,graphicx]{revtex4-1}
\makeatletter
\DeclareRobustCommand*\textsubscript[1]{%
  \@textsubscript{\selectfont#1}}
\def\@textsubscript#1{%
  {\m@th\ensuremath{_{\mbox{\fontsize\sf@size\z@#1}}}}}
\makeatother
\usepackage[latin1]{inputenc}
\usepackage{amsmath}
\usepackage{amsfonts}
\usepackage{amssymb}
\usepackage{graphicx,hyperref}
\usepackage{amssymb}
\usepackage{mathptmx}
\usepackage{listings}
\usepackage{slashbox}

\newcommand{\be}{\begin {equation}}

\newcommand{\tsp}{\textsuperscript}
\newcommand{\tsb}{\textsubscript}
\newcommand{\ee}{\end {equation}}
\newcommand{\beqa}{\begin {eqnarray}}
\newcommand{\eeqa}{\end {eqnarray}}
\newcommand{\mb}{\mathbf}
\hypersetup{
    colorlinks,%
    citecolor=blue,%
    filecolor=blue,%
    linkcolor=blue,%
    urlcolor=blue
}
\begin{document}


\title{Effects of cluster size and spatial laser intensity distribution on fusion neutron generation by laser driven Deuterium clusters} 


\author{Gaurav Mishra}
\email[]{gauravm@barc.gov.in}
\altaffiliation{Theoretical Physics Section, Bhabha Atomic Research Centre, Mumbai 400085, India}

\author{Amol R. Holkundkar}
\email[]{amol.holkundkar@pilani.bits-pilani.ac.in}
\altaffiliation{Department of Physics, Birla Institute of Technology and Science, Pilani 333 031, India}
 

\date{\today}

\begin{abstract}
A three dimensional molecular dynamic code is used to study the generation of fusion neutrons from Coulomb explosion of Deuterium clusters driven by intense near infra-red (NIR) laser ($\lambda=800nm$) of femtosecond pulse duration ($\tau=50 fs$) under beam-target interaction scheme. We have considered various clusters of average sizes ($\langle R_0\rangle$=80,140,200\AA) which are irradiated by a laser of peak spatial-temporal intensity  of 1$\times$10\tsp{18} W/cm\tsp{2}. The effects of cluster size and spatial laser intensity distribution on ion energies due to the Coulomb explosion of the cluster are included by convolution of single cluster single intensity ion energy distribution function (IEDF) over a range of cluster sizes and laser intensities. The final convoluted IEDF gets broadened on both lower and higher energy sides due to this procedure. Furthermore, the neutron yield which takes into account the convoluted IEDF, also gets modified by a factor of $\sim$2 compared to the case when convolution effects are ignored.
\end{abstract}


\maketitle 

\section{Introduction}
\label{sec:intro}The subject of interaction of high-intense laser with atomic clusters, acting as a bridge from gas to solid phase of matter, has witnessed a significant development on both experimental and theoretical ends during the last decade\cite{ditmire_detailed,lez,ditmire_MD,last2000_pra,Milchberg,jungreuthmayer2004_prl,Krainov2002,Fennel2010}. These clusters, generated by the isentropic expansion of a high pressure gas through a nozzle\cite{Hagena,Hagena1992}, exhibit nearly 100\% absorption of incident laser pulse energy\cite{ditmire_abs}. Near solid like density inside the cluster, plasma resonance and absence of any thermal dissipation effects were shown to be responsible for this significant absorption of laser energy. The consequence of this absorption is observed into the emission of highly-charged ions\cite{ditmire97}(with energies upto MeV\cite{dit_nature}), KeV electrons\cite{shao_electron} and x-rays ranging from Kev\cite{McPherson1994} to hundreds
of eV ($< 500$ eV)\cite{ditmire_xray,ditmire_detailed}. Laser-driven
particle accelaration\cite{fukuda}, coherent and incoherent X-ray
generation\cite{borisov}, nuclear fusion in deuterium clusters\cite{ditmire1999},
production of efficient plasma wave guides\cite{kumar_waveguide} and high orders
harmonic generation\cite{shim_harmonic} are few of the important
application of laser-cluster interaction

Investigation of Coulomb explosion of Deuterium (D) and Tritium (T) clusters is of prime importance due to the emission of high energy
(keV) D and T ions that can lead to generation of neutrons via the various nuclear fusion reactions (D + D  $\xrightarrow{\text{50\%}}$ He$^3$ + n, D + D $\xrightarrow{\text{50\%}}$ T + p, D + T $\rightarrow$ He$^4$ + n). The use of laser driven deuterium clusters as a neutron source is motivated by various factors like table-top dimensions of the whole set-up, nearly point like emission of neutrons from deuterium clusters, monochromaticity in the energy distribution, high-repetition rate and temporal durations as short as a few hundred picoseconds\cite{Zweiback2000_FusBurnTime}. The first direct demonstration of using laser irradiated clusters as neutron sources was shown by pioneering experiments performed by Ditmire and his group\cite{ditmire1999} with focusable laser intensity of 2$\times$10\tsp{16}W/cm\tsp{2} and average cluster size of nearly 50\AA.  These multi-KeV deuterium ions from neighbouring clusters undergoing Coulomb explosion fuse together to give neutrons of the characteristic energy of 2.45 MeV with a neutron yield of around $10^5$ fusion neutrons per joule of incident laser energy. In the experiments of Zweiback at al\cite{Zweiback2000_NucFus_PRL,Zweiback2002_FusDet_POP}, it was established that neutron yield was strongly dependent on the cluster size, laser focal geometry, and deuterium gas jet parameters. The role of laser pulse duration in the neutron yield was investigated by Madison \emph{et al}\cite{Madison2004_NeuPulDur}. The problem of nuclear fusion in laser-cluster interaction was also studied by Kishimoto \emph{et al} by using particle-in-cell methods\cite{Kishimoto2002}. It was observed that the expansion of the irradiated cluster was dependent on the two parameters : cluster size (a) and the electron excursion length {$\xi$} dependent upon the laser field (b). The two regimes of interest \emph {viz.} Coulomb explosion ($a\ll\xi$) and hydrodynamic ambipolar expansion ($a\gg\xi$) were identified. They found the high fusion neutron yield in Coulomb explosion regime with greater cluster size. Molecular dynamic studies performed by Last and Jortner \cite{Last2001_NeuHet_PRL,Last2001_NeuHet_PRA} found enhanced neutron yield from the Coulomb explosion of heteronuclear clusters $(D_2O)_n$, as compared with homonuclear clusters $(D)_n$. The increased neutron yield for $(D_2O)_n$ clusters was the result of higher kinetic energies of $D^+$ ions triggered by the highly charged $O^{+q}$ ions. This theoretical finding was further supported by experiments conducted by Madison \emph{et al}\cite{Madison2004_FusHet_POP}. By using molecular dynamic model, Petrov and Davis \cite{Davis2006_NeuYie_POP,Petrov2006_NeuYie_PPCF} studied the neutron production from high intensity laser-cluster interaction in an alternate beam-target interaction scheme. They used the laser driven clusters as a source of high energy deuterium ions which reacted with the walls of a surrounding fusion reaction chamber with walls coated with DT fuel or other deuterated material such as $CD_2$ and generated a large amount of neutrons. They estimated a high neutron yield of $10^6-10^7$ neutrons/Joule with the laser intensity of $10^{16}-10^{18}W/cm^2$ and clusters with initial radius of $~20nm$.

All of these studies indicate the strong dependence of neutron yield on the distribution of D ion energies as a outcome of Coulomb explosion of D clusters. Complete understanding of experimentally observed IEDF requires the inclusion of cluster size\cite{Mendham2001} and spatial laser intensity distribution\cite{LCI_Posthu} effects in the analytical/computational single cluster single intensity IEDF\cite{Islam,Heidenreich}. The analytical studies carried out by Islam \emph{et al.} \cite{Islam} take into account the effect of long-normal cluster size distribution and two dimensional (2-D) spatial laser intensity variation\cite{Heidenreich} in the complete CVI approximation\cite{LastCVI}. The more general computational approach used by Heidenreich \emph{et al.} first calculates the single cluster single intensity IEDF and then makes the use of double averaging over cluster size and 2-D/3-D(three dimensional) spatial laser intensity profile to determine final doubly averaged IEDF. It is important to note that this approach is more general than the one used by Islam \emph{et al.} as it does not require the complete CVI condition. We may also note that the effect of attenuation of laser intensity on fusion neutron yield while propagating through an assembly of much larger clusters termed as nano-droplets has been investigated by Ron \emph{et al.}\cite{RonDroplet}. Their results have indicated that large nanodroplets (size greater than 140 nm) are sensitive towards strong laser intensity attenuation that significantly affects the resulting neutron yield. The combined effect of laser intensity attenuation and cluster size distribution on fusion neutron yield is investigated in the framework of modified Coulomb explosion model\cite{LiSizeIntenAtten}.	

Earlier, we developed a molecular dynamic (MD) code - MDILAC (serial\cite{HolMDSer} as well as parallel on openMP framework\cite{HolMDPar}) to study the interaction dynamics of medium and large sized atomic clusters driven by intense femtosecond laser fields\cite{Mishra2011,MishraAniso_2011,MishraAniso_2012,MishraCEP_2013}. In this paper, we present our computational studies on the fusion neutron yield from D clusters under beam-target interaction scheme design by taking into account the effects of cluster size and spatial laser intensity distribution. For these studies, ion energy distribution function (IEDF) obtained from MDILAC is convoluted over cluster size and laser intensity as suggested by Heidenreich \emph{et al.} \cite{Heidenreich} which is further used to calculate the neutron yield. The details of computational scheme is given in section \ref{sec:simu_meth}. The results are discussed in section \ref{sec:res}. Finally we conclude the paper in section \ref{sec:conc}. 

\section{Simulation methodology}
\label{sec:simu_meth}In this section, we intend to give the simulation scheme employed to calculate the neutron yield from laser driven deuterium clusters in the beam-target interaction scheme\cite{Davis2006_NeuYie_POP,Petrov2006_NeuYie_PPCF}. The single cluster single intensity ion energy distribution function (IEDF) of D ions is determined by the molecular dynamic code $-$ MDILAC\cite{HolMDSer,HolMDPar}. In this code, a spherical cluster of size (R\tsb{0}) is considered to be at the center of a three dimensional simulation box which is irradiated by the a high intensity Gaussian (time as well as space) laser pulse of near infra-red wavelength ($\lambda =800$nm). The corresponding number of atoms can be calculated as 
\be
N=\frac{R_0}{R_W},
\label{eq:n_atoms}
\ee
where R\tsb{W} is the Weigner-Seitz radius of the constituent atom (R\tsb{W} = 1.7\AA\ for D).
The spatial variation of laser pulse is ignored as the size of cluster is small enough compared to the wavelength of the laser. The ionization of the cluster atoms is due to both tunnel ionization\cite{ammosov1986} and collisional ionization\cite{lot,vor}. Charged particles inside the cluster move in the presence of laser electromagnetic field and Coulomb field of the other charged particles. The phase space of the charged particles is stored by solving the relativistic equation of motion ($d\mb{p_i}/dt=\mb{F_i}$, $\mb{v_i}=\mb{p_i}/(m_i\sqrt{1 + |\mb{p_i}|^2/(m_i c)^2})$, $d\mb{r_i}/dt=\mb{v_i}$; where $\mb{p_i}$, $\mb{v_i}$, $\mb{r_i}$ and $m_i$ are relativistic momentum, velocity, coordinate and mass of the $i^{th}$ particle.) which is further used to determine the IEDF.

The single cluster single intensity IEDF (F)  obtained from MDILAC is convoluted over cluster size and spatial laser intensity distribution\cite{Heidenreich} to obtain the final convoluted IEDF (F\tsb{c}) as,
\be
F_c=\frac{\int\limits_{I_{mn}}^{I_{mx}}\int\limits_{n_{mn}}^{n_{mx}}n\ \xi(I)\: \chi(n)\:F\ dn\ dI\:}{\int\limits_{I_{mn}}^{I_{mx}}\int\limits_{n_{mn}}^{n_{mx}}n\ \xi(I)\: \chi(n)\: dn\ dI\:},
\label{eq:iedf_conv}
\ee
where $\xi(I)$ represents the laser intensity distribution function with I\tsb{mn} and I\tsb{mx} as minimum and maximum laser intensity and $\chi(n)$ designates the cluster size (in terms of number of particles per cluster) distribution with n\tsb{mn} and n\tsb{mx} as minimum and maximum cluster size. The spatial intensity distribution function $\xi(I)$ for a 2-D Gaussian laser pulse\cite{LCI_Posthu} ($I(r)=I_p \exp(-2r^2/w_0^2)$ with I\tsb{p} as spatially peak laser intensity and w\tsb{0} as minimum beam waist radius) can be written as
\be
\xi(I)=\frac{1}{I}\:\left[\log\left(\frac{I_{mx}}{I_{mn}}\right)\right]^{-1}.
\label{eq:dis_inten}
\ee It has been shown experimentally\cite{Mendham2001} that clusters follow a long-normal distribution in the interaction regime which is given as, 
\be
\chi(n)=\frac{1}{n\sqrt{2\pi\sigma^2}}\exp\left[-\frac{\left\lbrace\log(n)-\mu\right\rbrace^2}{2\sigma^2}\right],
\label{eq:dis_size}
\ee
where $\mu$ and average cluster size $\left\langle n \right\rangle$ are related as, $\left\langle n \right\rangle=\exp\left(\mu+\sigma^2/2\right)$, and $\sigma=0.4087$.

Now, D ions after the Coulomb explosion of deuterium cluster traverses through D or T coated surrounding material where there create the neutrons as a result of D-D or D-T nuclear fusion reaction. The fusion neutron yield in such a beam-target interaction scheme can be calculated as\cite{Hol_Neu},
\be
Y=\left\langle y \right\rangle N_d,
\label{eq:neu_yield}
\ee where $\left\langle y \right\rangle$ is the average neutron yield per ion and $N_d$ is the total number of D ions produced as a result of Coulomb explosion. The average neutron yield per ion ($\left\langle y \right\rangle$) is defined as 
\be
\left\langle y \right\rangle=\int_{0}^{E^{mx}}F_c\:y(E)dE,
\label{eq:ave_neu_yield}
\ee
where F\tsb{c} is the convoluted IEDF given by Eq.(\ref{eq:iedf_conv}), E\tsb{mx} is the maximum energy of D ion after convolution. y(E) appearing in Eq.(\ref{eq:ave_neu_yield}) is the fusion neutron yield for a D ion with initial energy E which is determined as,
\be
y(E)=\int_{0}^{E}\sigma(\epsilon)/S(\epsilon)\:d\epsilon
\label{eq:neu_yield_ion}
\ee. 
Here, $\sigma$ is the fusion cross section taken from \emph{Huba}\cite{Huba} and $S(\epsilon)$ is the stopping power of D ions normalised over the target density of $5\times10^{22}$cm$^{-3}$ which is determined by SRIM\cite{Ziegler2010}. It is important to note that complete determination of fusion yield in Eq.(\ref{eq:neu_yield}) also requires the number of D ions (N\tsb{d}) produced per unit joule of laser energy absorbed as a outcome of Coulomb explosion of D cluster. These are the ions which further interacts with the target material to produce neutrons. Keeping the experimental observation in mind that laser driven clusters almost absorb all incident laser energy upon them\cite{ditmire_abs} and ignoring the effect of laser intensity attenuation in medium sizes clusters\cite{RonDroplet}, one cane denote the $\eta$ as the fraction of absorbed energy transferred to ions (conversion efficiency of laser energy to ions energy). The total number of D ions per unit of laser energy absorbed can be estimated as,
\be
N_d=\eta E_{ls}/E_{av},
\label{eq:n_ions}
\ee
where E\tsb{ls} is the laser pulse energy and E\tsb{av} is the average kinetic energy of ions.
\section{Results and discussion}
\label{sec:res}
\subsection{Energetics of cluster Coulomb explosion}
We have calculated the neutron yield for various average cluster sizes by incorporating the effects of cluster size and spatial laser intensity distribution on the corresponding single cluster single intensity IEDF. The single cluster single intensity IEDF is determined by studying the interaction dynamics of various D clusters of average sizes ($\left(\left\langle R \right\rangle=80,140,200\AA\right)$) driven by near-infrared laser (NIR, $\lambda=800$nm) with FWHM pulse duration of 50 fs and peak spatial-temporal intensity (I\tsb{p}) of $10^{18}W/cm^2$. The corresponding average number of atoms for different average cluster size can be calculated from Eq.\ref{eq:n_atoms} to give the values as $\left\langle n \right\rangle=1.04\times10^5,5.58\times10^5,1.62\times10^6$. The different values of cluster size for each average cluster size used for size convolution can be determined from the long normal distribution function given by Eq.(\ref{eq:dis_size}) and are shown in Table \ref{tab:table1}.
\begin{table}
\caption{\label{tab:table1}Single cluster sizes (R\tsb{1}-R\tsb{6}) used for size convolution of different average cluster radii (\textless R\tsb0\textgreater) }
\begin{ruledtabular}
\begin{tabular}{ccccccc}
 $\left\langle R_0 \right\rangle$ (\AA) & $R_1$ (\AA) & $R_2$ (\AA)& $R_3$ (\AA) & $R_4$ (\AA) & $R_5$ (\AA) & $R_6$ (\AA) \\
\hline
80 & 52 & 62 & 73 & 82 & 90 & 107\\
140 & 90 & 110 & 128 & 144 & 158 & 188\\
200 & 135 & 168 & 190 & 210 & 225 & 270
\label{tab:dis_size}
\end{tabular}
\end{ruledtabular}
\end{table}
The effect of intensity convolution can be accounted by taking the different values of intensities as $1\times 10^{18},8\times 10^{17},4\times 10^{17},1\times 10^{17}$W/cm$^2$ obtained from spatial laser intensity distribution function given by Eq.(\ref{eq:dis_inten}) for peak value of intensity $1\times 10^{18}$W/cm$^2$ used in the studies. 

\begin{figure}
\centering
 \includegraphics[height=3.5in,width=5.0in]{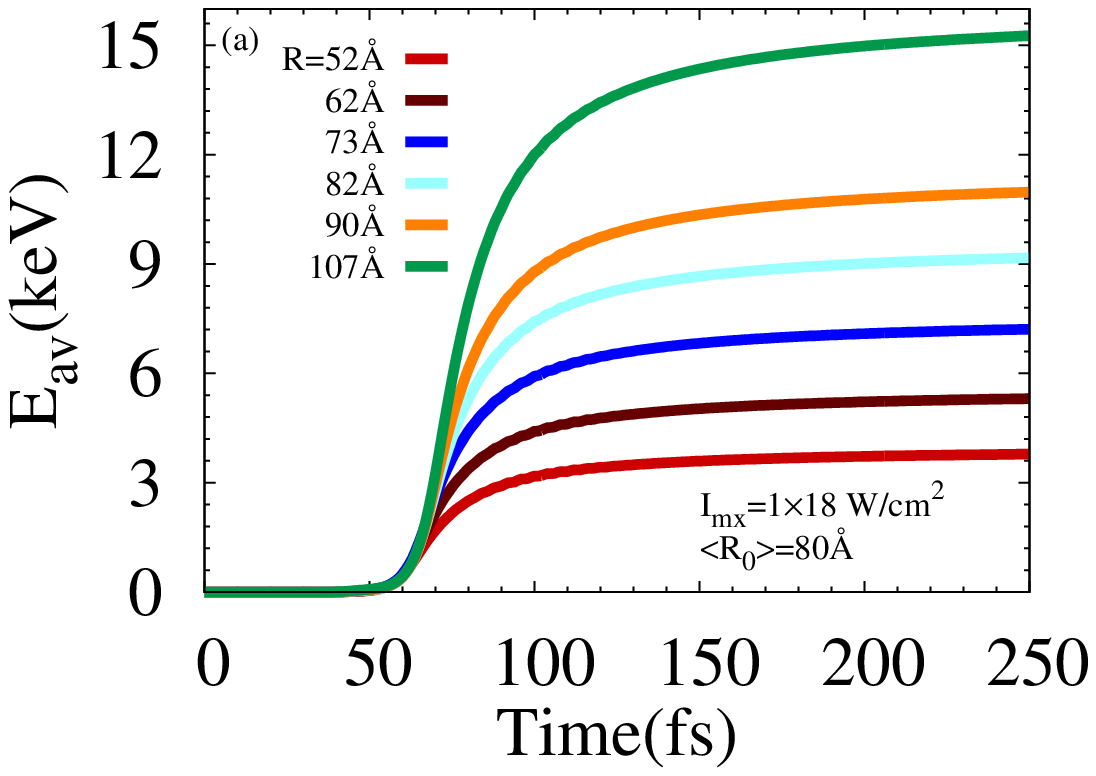}
\hspace{-.2in}
\vspace{0 in}
\includegraphics[height=3.5in,width=5.0in]{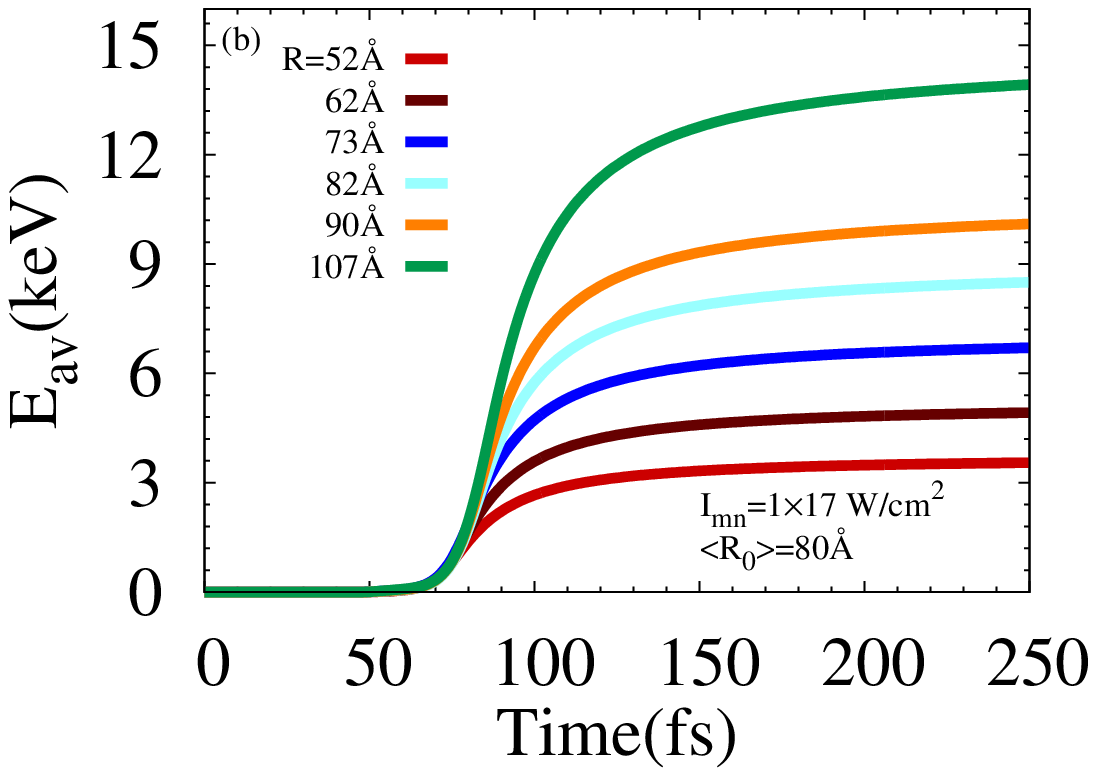}
 \caption{(color on-line) Temporal variation of average kinetic energy of ions from Coulomb explosion of deuterium clusters of various cluster sizes (52 \AA\ - 107 \AA) at laser intensity I\tsb{mx} of 1$\times$10\tsp{18} W/cm\tsp{2} (a) and I\tsb{mn} of 1$\times$ 10\tsp{17} W/cm\tsp{2} (b). The average cluster size for these simulations is $\langle R_0\rangle=80$ \AA\ and pulse duration of the laser is 50 fs. The peak spatial-temporal intensity of the laser for these simulations is taken as 1$\times$10\tsp{18} W/cm\tsp{2}.}
 \label{fig.1}
\vskip 0.5in
\hskip 4.5in G. Mishra et \emph{al.}, Fig. 1
 \end{figure}

\begin{figure}
\centering
 \includegraphics[height=3.5in,width=5.0in]{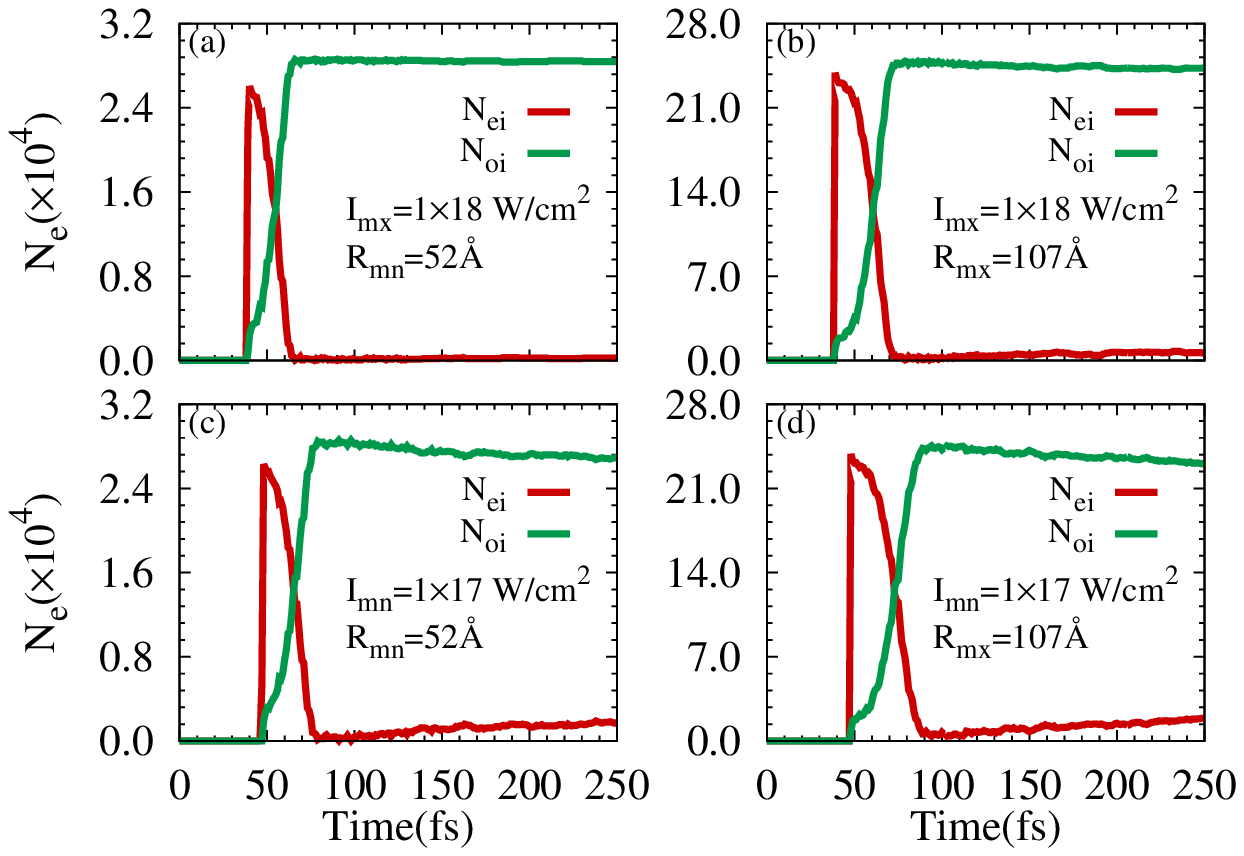}
\caption{(color on-line) Time variation of inner and outer electron population for I\textsubscript{mx}=1$\times$10\tsp{18} W/cm\tsp{2}, R\tsb{mn}=52\AA\ (a) I\tsb{mx}=1$\times$10\tsp{18} W/cm\tsp{2}, R\tsb{mn}=107\AA\ (b) I\tsb{mx}=1$\times$10\tsp{17} W/cm\tsp{2}, R\tsb{mn}=52\AA\ (c) and I\tsb{mx}=1$\times$10\tsp{17} W/cm\tsp{2}, R\tsb{mn}=107\AA\ (d). The other laser and cluster parameters are same as used in Fig.\ref{fig.1}.}
 \label{fig.2}
\vskip 0.5in
\hskip 4.5in G. Mishra et \emph{al.}, Fig. 2
 \end{figure}

Now we discuss the simulation results for interaction dynamics of D cluster of average size 80\AA\ driven by laser of peak spatial-temporal intensity of 1$\times$10\tsp{18} W/cm\tsp{2}. For this case, we have carried out the single cluster single intensity MD simulations for cluster size range (R\tsb{mn}=52\AA\ - R\tsb{mx}=107\AA) and laser intensity range (I\tsb{mx}=1$\times$10\tsp{18} W/cm\tsp{2} to I\tsb{mn}=1$\times$10\tsp{17} W/cm\tsp{2}) as mentioned in the above paragraph. In Fig.\ref{fig.1}(a) and (b), we present the plots of temporal variation of average ion kinetic energy (E\tsb{av}) for various cluster sizes (R\tsb{mn} - R\tsb{mx}) at maximum intensity I\tsb{mx} and minimum intensity I\tsb{mn}. For each cluster radius, E\tsb{av} remains zero for certain time, then it increases quickly and finally, it saturates. This time dependent behaviour of E\tsb{av} is due to the Coulomb explosion of cluster as a result of time dependent inner and outer ionization\cite{LastInOut} of the cluster as shown in Fig.\ref{fig.2}. Initially, the laser intensity is not sufficiently enough to create any inner or outer ionization. After some time, the inner ionization of cluster atoms commences at the appropriate value of the incident time dependent laser intensity which is followed by the slow outer ionization of the cluster. Finally the cluster undergoes Coulomb explosion as a result of positive charge build-up inside the cluster due to the outer ionization of the cluster. This leads to the rapid rise in E\tsb{av} and finally, it saturates. 

\begin{figure}
\centering
 \includegraphics[height=3.5in,width=5.0in]{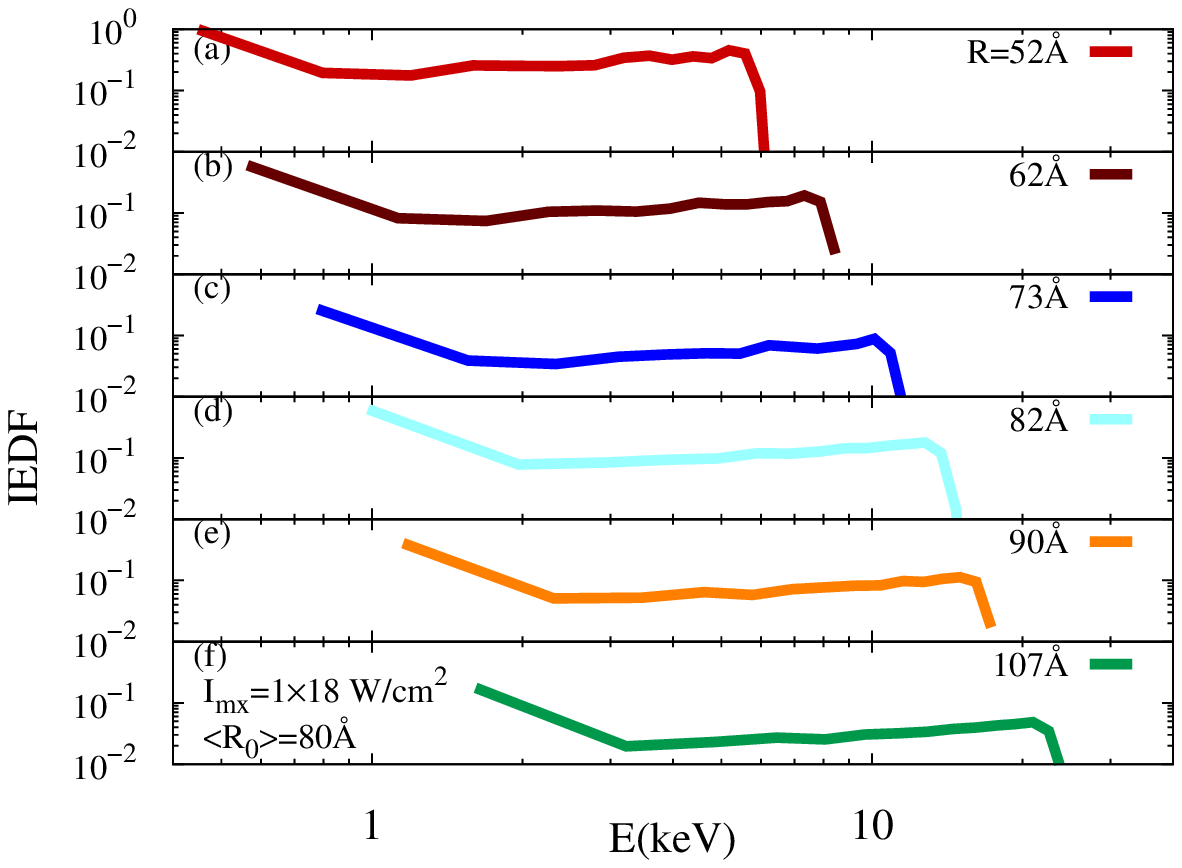}
\hspace{-.2in}
\vspace{0 in}
\includegraphics[height=3.5in,width=5.0in]{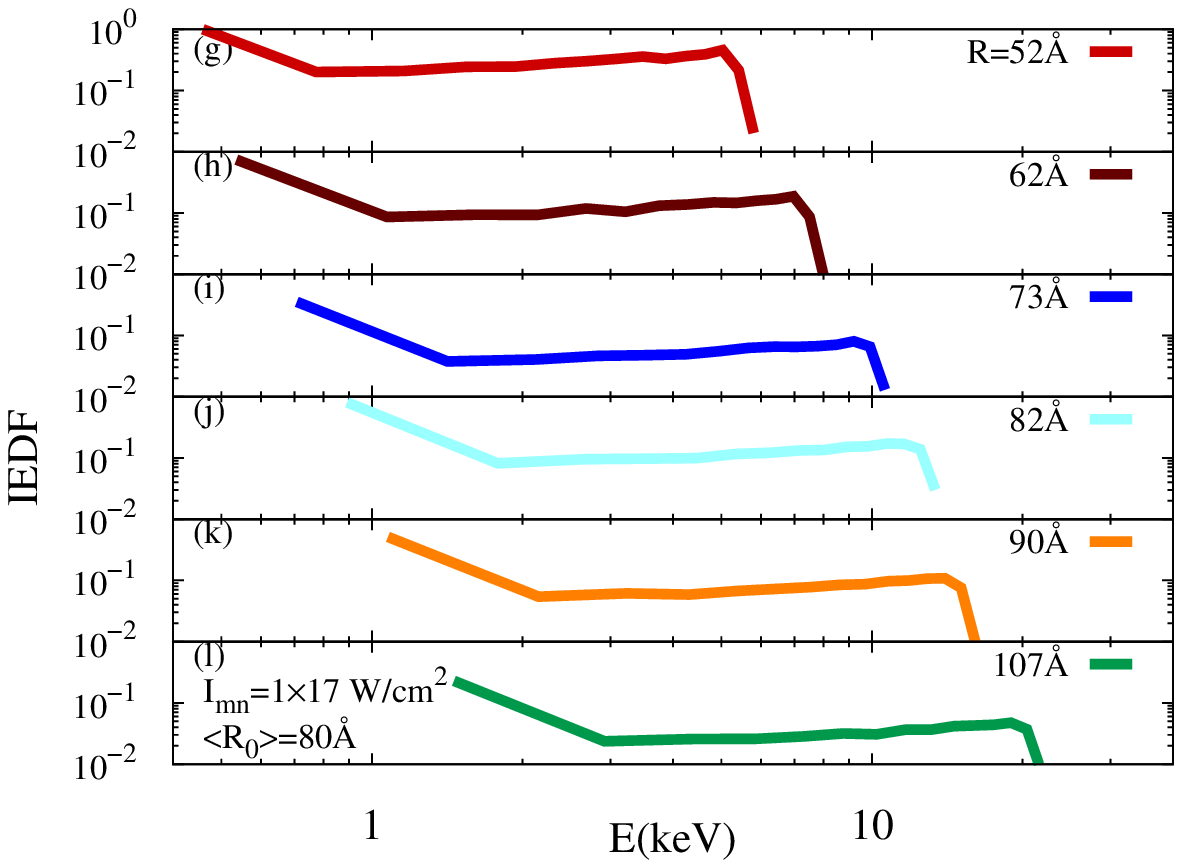}
 \caption{(color on-line) Single cluster IEDF for various cluster sizes (52 \AA\ - 107 \AA) at laser intensity I\tsb{mx} of 1$\times$10\tsp{18} W/cm\tsp{2} (a-f) and I\tsb{mn} of 1$\times$ 10\tsp{17} W/cm\tsp{2} (g-l). The other laser and cluster parameters are same as used in Fig.\ref{fig.1}.}
 \label{fig.3}
\vskip 0.5in
\hskip 4.5in G. Mishra et \emph{al.}, Fig. 3
 \end{figure}

It is also important to see that as the size of the cluster increases from R\tsb{mn} to R\tsb{mx}, the occurrence of Coulomb explosion also gets delayed. This can be explained by observing the Figs.\ref{fig.2}(a) and (b). These plots show that the timing of the inner electron population rise remains unaffected with the increase in the cluster radius as the process of inner ionization depends upon the strength of incident laser field or laser intensity. On the other hand, outer ionization is getting delayed with the increase in the cluster size at fixed laser intensity. This is due to the fact that the number of D ions as a result of inner ionization is high for large clusters which makes outer ionization difficult. Consequently, a higher electric field strength is required to facilitate the process of outer ionization which becomes possible at later times in the laser intensity envelope. The plots for single cluster IEDF are shown in Fig.\ref{fig.3} for various cluster sizes (R\tsb{mn} - R\tsb{mx}) at maximum intensity I\tsb{mx} and minimum intensity I\tsb{mn}. For comparison, we have shown the saturated values of average kinetic energy (E\tsb{av}) and maximum value of ion kinetic energy (E\tsb{mx}) in Tab.\ref{tab:table2}
\begin{table}
\caption{\label{tab:table2}Saturated values of average (E\tsb{av}) and maximum ion kinetic energy (E\tsb{mx}) for complete cluster size range (R\tsb{mn}-R\tsb{mx}) at maximum I\tsb{mx} and minimum laser intensities I\tsb{mn}. The average radius of the cluster is taken as 80\AA\ whereas the peak spatial-temporal intensity of laser is kept at 1$\times$10\tsp{18} W/cm\tsp{2}. }
\begin{ruledtabular}
\begin{tabular}{ccccc}
 &\multicolumn{2}{c}{I\tsb{mx}=1$\times$10\tsp{18} W/cm\tsp{2}}&\multicolumn{2}{c}{I\tsb{mn}=1$\times$10\tsp{17} W/cm\tsp{2}}\\
 R(\AA)&E\tsb{av}(KeV)&E\tsb{mx}(KeV)&E\tsb{av}(KeV)
&E\tsb{mx}(KeV)\\ \hline
 52 & 3.78 & 6.38 & 3.55 & 5.80 \\
 62 & 5.31 & 8.45 & 4.92 & 8.57\\
 73 & 7.21 & 12.45 & 6.70 & 10.62\\
 82 & 9.17 & 15.72 & 8.50 & 13.39\\
 90 & 10.97 & 17.32 & 10.10 & 17.23\\
 107 & 15.26 & 24.20 & 13.92 & 23.34\\
 \end{tabular}
\end{ruledtabular}
\end{table}
for these cluster radii and laser intensities. We may note that E\tsb{av} and E\tsb{mx} are 3.78 KeV and 6.38 KeV for R\tsb{mn} which increase up to 15.26 KeV and 24.20 KeV for R\tsb{mx} at maximum laser intensity I\tsb{mx}. 

\begin{figure}
\centering
 \includegraphics[height=3.5in,width=5.0in]{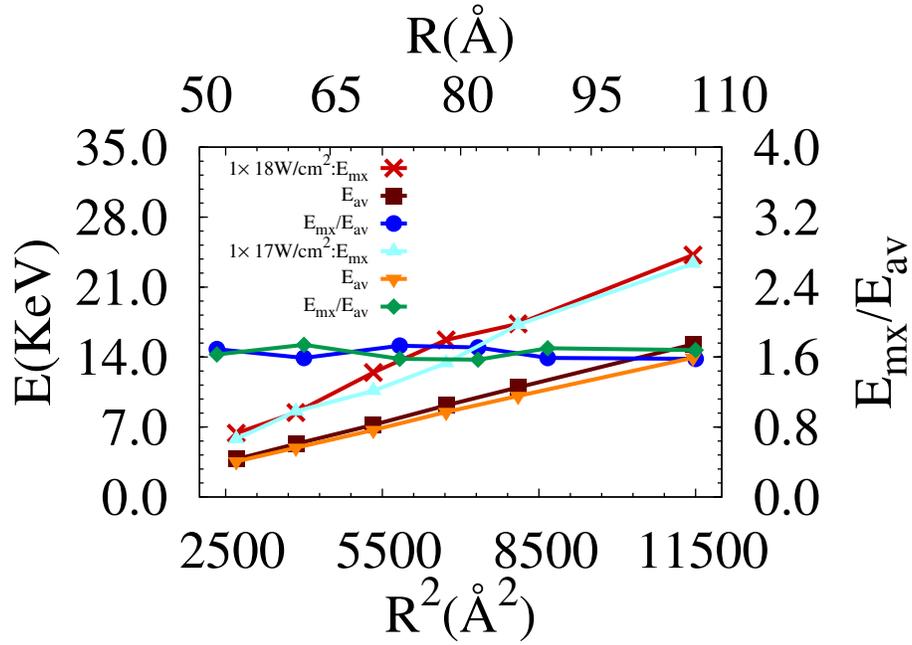}
\caption{(color on-line) Variation of maximum (E\tsb{mx}), average ion kinetic energy (E\tsb{av}) and ratio of maximum to average ion kinetic energy (E\tsb{mx}/E\tsb{av}) with cluster size. The laser and cluster parameters are same as used in Fig.\ref{fig.1}.}
 \label{fig.4}
\vskip 0.5in
\hskip 4.5in G. Mishra et \emph{al.}, Fig. 4
 \end{figure}

\begin{figure}
\centering
 \includegraphics[height=3.5in,width=5.0in]{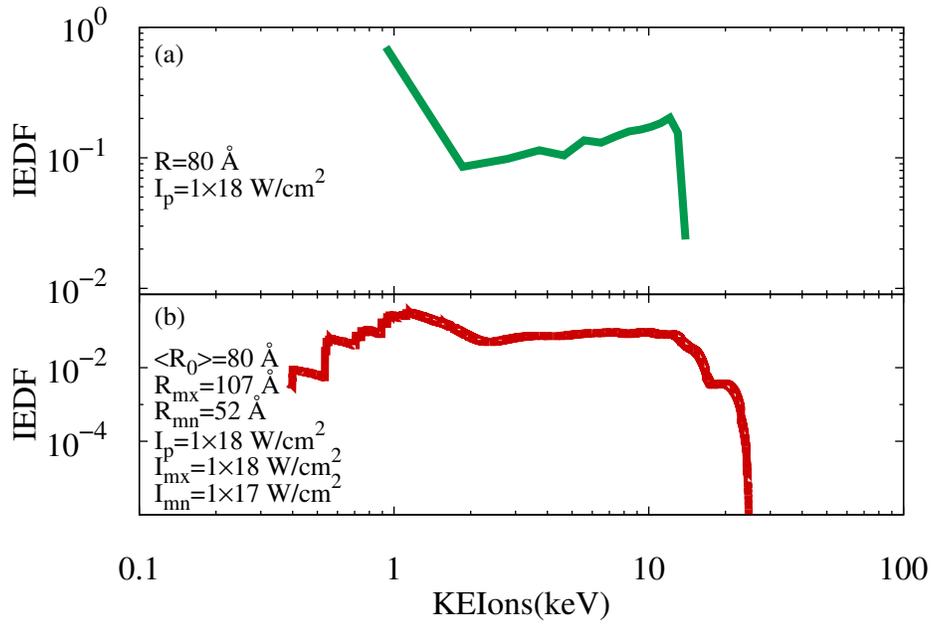}
\caption{(color on-line) IEDF without convolution (a) and with convolution over cluster size and laser intensity (b) for average cluster radius $\langle R_0\rangle=80$ \AA\ . The peak spatial-temporal intensity of the laser is taken as I\tsb{p}=1$\times$10\tsp{18} W/cm\tsp{2} and FWHM pulse duration of the laser is 50 fs.}
 \label{fig.5}
\vskip 0.5in
\hskip 4.5in G. Mishra et \emph{al.}, Fig. 5
 \end{figure}

In Fig.\ref{fig.4},we have also shown the variation of E\tsb{av},E\tsb{mx} and E\tsb{mx}/E\tsb{av} as a function of cluster radius for both the extreme intensity values. It is important to note from Fig.\ref{fig.4} that both E\tsb{av} and E\tsb{mx} goes as R\tsp{2} and ratio of E\tsb{mx} to E\tsb{av} remains nearly close to 5/3 for extreme intensity values (I\tsb{mx} and I\tsb{mn}) used in the simulation. All of these characteristics demonstrate the energetics of Coulomb explosion under single cluster approximation. As the size of the cluster increases at high laser intensity, the residual cluster charge build-up is more for larger clusters due to enhanced outer ionization than that for smaller clusters (Fig.\ref{fig.2}(a) and (b)). Consequently, the strength of Coulomb explosion is more for large clusters that leads to higher kinetic energy of ions (E\tsb{av} as well as E\tsb{mx}) for large clusters than that for small clusters. For the case of minimum laser intensity I\tsb{mn} (Figs.\ref{fig.1}(b) and \ref{fig.3}(g-l)), the trend for temporal variation of E\tsb{av} and single cluster IEDF remains similar to that (Figs.\ref{fig.1}(a) and \ref{fig.3}(a-f)) observed at highest laser intensity I\tsb{mx} for complete  cluster size range (R\tsb{mx}-R\tsb{mn}). The only difference is that the saturated values of time dependent E\tsb{av} and maximum ion kinetic energy E\tsb{mx} are slightly smaller than those values at maximum laser intensity I\tsb{mx} as shown in Tab.\ref{tab:table2}. This slight  difference is attributed to the small values of saturated inner electron population (Fig.\ref{fig.2}(c) and (d)) at lower intensity which inhibit the strength of Coulomb explosion and further reduce the average and maximum ion kinetic energies. We note that the energetics of Coulomb explosion of D cluster is strongly dependent on the cluster size and weakly dependent on laser intensity for the average cluster size of 80\AA\ and peak spatial-temporal intensity of 1$\times$10\tsp{18}. The above discussion reflects the use of both size and intensity convolution to calculate the final convoluted IEDF (F\tsb{c}) which will be further used to determine the neutron yield. In Figs.\ref{fig.5}(a) and (b), we present the IEDF without convolution and with convolution, respectively for this set of laser and cluster parameters. We may note that energy spectrum has broadened on both sides of energy values after convolution. The minimum and maximum ion kinetic energies after Coulomb explosion are 847 ev and 12.70 Kev without taking into account the effect of spatial laser intensity profile and cluster size distribution. The minimum ion kinetic energy reduces to 387 ev whereas the maximum ion kinetic energy enhances up to 25.30 KeV after convolution. 

\begin{figure}
\centering
 \includegraphics[height=3.5in,width=5.0in]{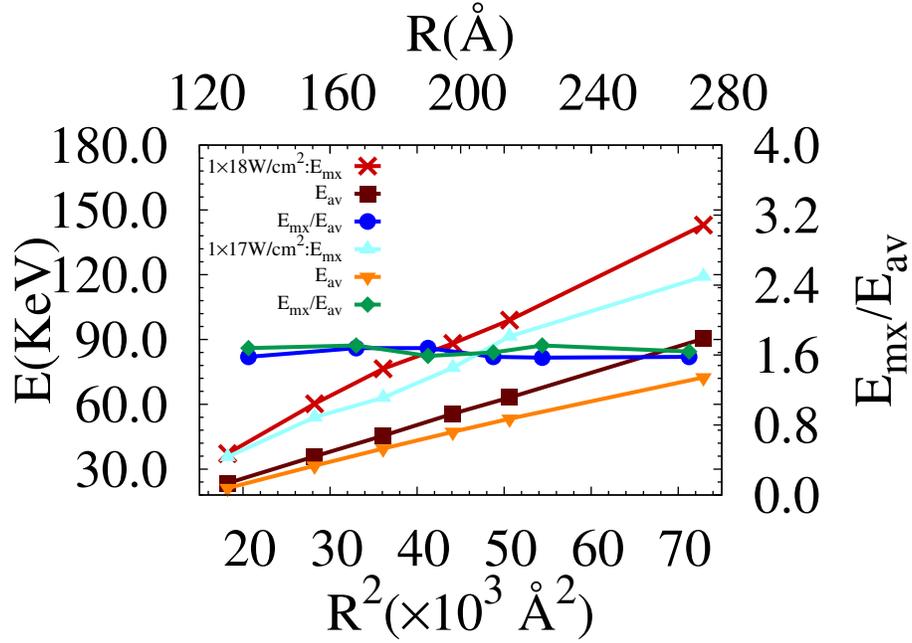}
\caption{(color on-line) Variation of maximum (E\tsb{mx}) ,average ion kinetic energy (E\tsb{av}) and ratio of maximum to average ion kinetic energy (E\tsb{mx}/E\tsb{av}) with cluster size. The average cluster size for these simulations is $\langle R_0\rangle=200$ \AA\ and pulse duration of the laser is 50 fs. The peak spatial-temporal intensity of the laser for these simulations is taken as 1$\times$10\tsp{18} W/cm\tsp{2}.}
 \label{fig.6}
\vskip 0.5in
\hskip 4.5in G. Mishra et \emph{al.}, Fig. 6
 \end{figure}

\begin{figure}
\centering
 \includegraphics[height=3.5in,width=5.0in]{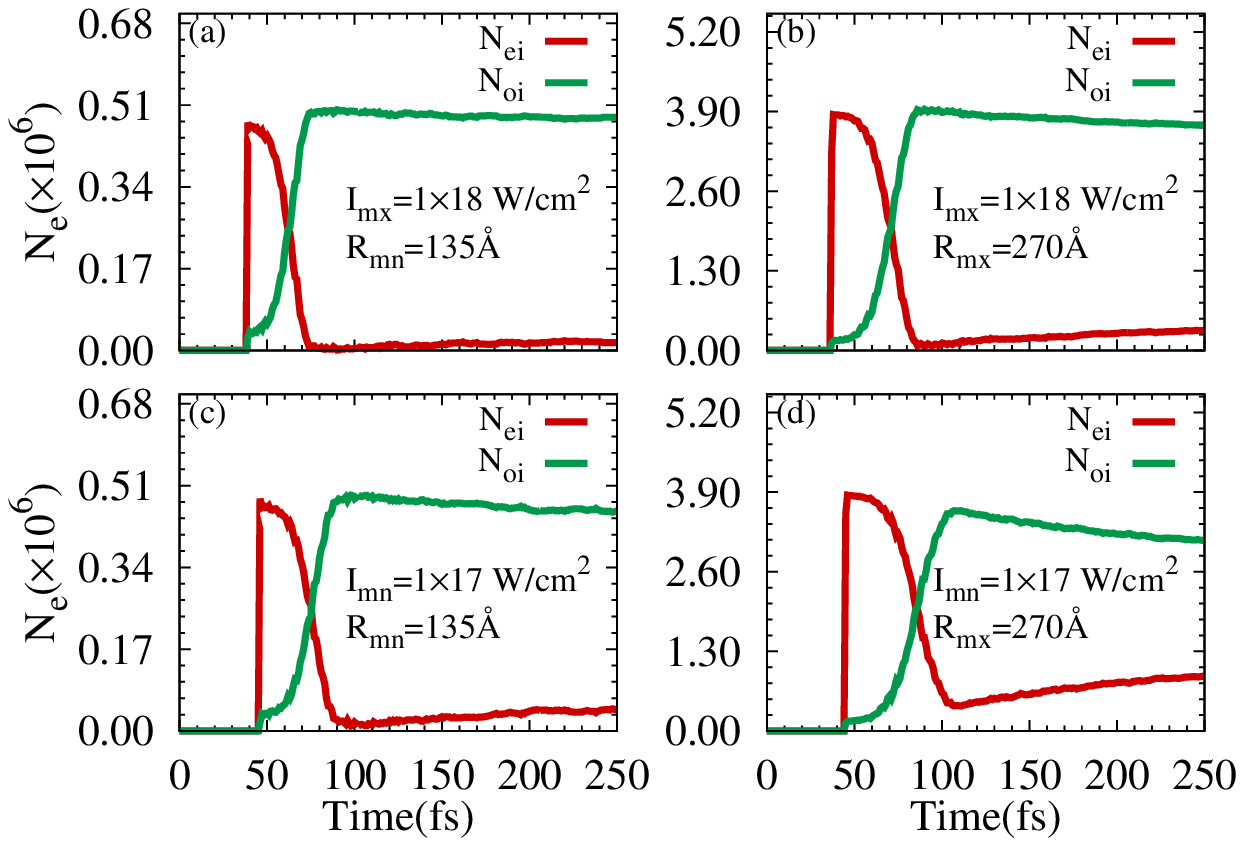}
\caption{(color on-line) Time variation of inner and outer electron population for I\textsubscript{mx}=1$\times$10\tsp{18} W/cm\tsp{2}, R\tsb{mn}=135\AA\ (a) I\tsb{mx}=1$\times$10\tsp{18} W/cm\tsp{2}, R\tsb{mn}=270\AA\ (b) I\tsb{mx}=1$\times$10\tsp{17} W/cm\tsp{2}, R\tsb{mn}=135\AA\ (c) and I\tsb{mx}=1$\times$10\tsp{17} W/cm\tsp{2}, R\tsb{mn}=270\AA\ (d). The other laser and cluster parameters are same as used in Fig.\ref{fig.6}.}
 \label{fig.7}
\vskip 0.5in
\hskip 4.5in G. Mishra et \emph{al.}, Fig. 7
 \end{figure}

\begin{figure}
\centering
 \includegraphics[height=3.5in,width=5.0in]{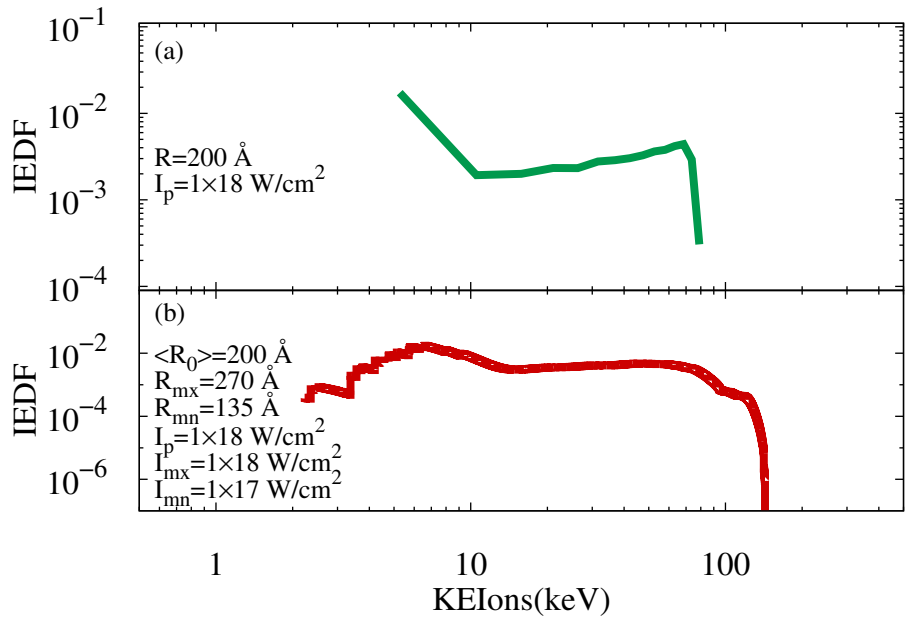}
\caption{(color on-line) IEDF without convolution (a) and with convolution over cluster size and laser intensity (b) for average cluster radius $\langle R_0\rangle=200$ \AA\ and peak spatial-temporal laser intensity of I\tsb{p}=1$\times$10\tsp{18} W/cm\tsp{2}. The FWHM pulse duration of the laser is 50 fs.}
 \label{fig.8}
\vskip 0.5in
\hskip 4.5in G. Mishra et \emph{al.}, Fig. 8
 \end{figure}

Next we examine the energetics of the largest cluster used in our studies of average size of 200\AA\ which is irradiated by peak spatial-temporal intensity of 1$\times$10\tsp{18} W/cm\tsp{2}. For this case, we have carried out the single cluster single intensity MD simulations for cluster size range (R\tsb{mn}=135\AA\ - R\tsb{mx}=270\AA) as mentioned in Table\ref{tab:dis_size}. The different laser intensities used for intensity convolution lie in the same range (I\tsb{mx}=1$\times$10\tsp{18} W/cm\tsp{2} to I\tsb{mn}=1$\times$10\tsp{17} W/cm\tsp{2}) as the peak spatial-temporal intensity is identical to what used previously for irradiation of cluster of average size  80\AA. The results for E\tsb{av}, E\tsb{mx} and E\tsb{mx}/E\tsb{av} for complete cluster size range (R\tsb{mn}-R\tsb{mx}) at two extreme intensity values I\tsb{mx} and I\tsb{mn} are presented in Fig.\ref{fig.6}. For each intensity, both of the energy parameters E\tsb{mn} and E\tsb{mx} goes as square of the cluster radius and ratio E\tsb{mx}/E\tsb{mn} is close to 5/3. The difference between the two energy values at two extreme intensity points increases significantly as the cluster size is increased. Moreover, this difference is larger than that observed for the average cluster size of $\langle R_0\rangle=80$ \AA\ (Fig.\ref{fig.4}). This can be explained by the time dependent population of inner (N\tsb{ei}) and outer electrons (N\tsb{eo}) for the case of two extreme cluster radii (R\tsb{mn} and R\tsb{mx}) and intensities (I\tsb{mn} and I\tsb{mx}) as shown in Fig. \ref{fig.7}. We note from Figs. \ref{fig.7}(b) and (d) that ratio of inner electron population to outer electron population level (N\tsb{ei}/N\tsb{eo}) increases from 9.89 \% to 28.71\% for R\tsb{mx} when the intensity is lowered from I\tsb{mx} to I\tsb{mn}. The enhanced inner electron population at I\tsb{mn} lowers the strength of Coulomb explosion which leads to lowered energy values. This discussion reflects that the energetics of Coulomb explosion for large clusters ($\langle R_0\rangle=200$ \AA) is strongly dependent on both cluster sizes and laser intensities used for convolution. The significance of intensity convolution is much more for this case than that for smaller clusters ($\langle R_0\rangle=80$ \AA). The convoluted IEDF along with IEDF without convolution are presented in Fig.\ref{fig.8} for the case of $\langle R_0\rangle=200$ \AA\ which shows the the broadening of energy spectrum at both extreme energy points compared to the IEDF without convolution. 

\begin{figure}
\centering
 \includegraphics[height=3.5in,width=5.0in]{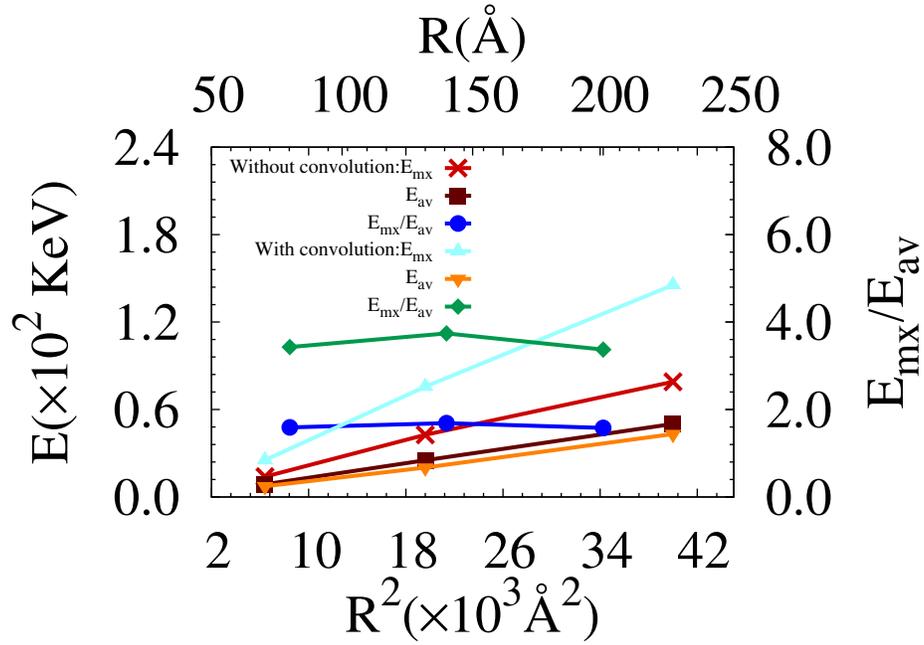}
\caption{(color on-line) Variation of maximum (E\tsb{mx}) and average ion kinetic energy (E\tsb{av}) as a function of average cluster radius. The other simulation parameters are same as used in Fig.\ref{fig.8}.}
 \label{fig.9}
\vskip 0.5in
\hskip 4.5in G. Mishra et \emph{al.}, Fig. 9
 \end{figure}

To show the effect of convolution on various energy parameters, we have shown E\tsb{mx} and E\tsb{av} in the Fig.\ref{fig.9} as a function of square of average cluster radius and ratio of maximum energy to average energy (E\tsb{mx}/E\tsb{av}) as a function of average cluster radius . It is important to note that E\tsb{mx} increases significantly after convolution for higher average cluster size whereas E\tsb{av} slightly reduces after convolution. This is due to the fact that cluster sizes considered for size averaging follow a long normal distribution and a significant contribution in maximum ion energies comes from larger clusters used in the size convolution. On the other hand, laser intensity distribution function goes as inverse of intensity and maximum contribution comes from the highest intensity used in the simulation. All other intensity points are lower than that this value and they affect only on the lower and medium side of the convoluted energy spectrum. It is also important to note that the ratio of maximum energy to average energy (E\tsb{mx}/E\tsb{av}) modifies to 3.51 after convolution from 1.62 for the case where convolution is not considered.

\subsection{Fusion neutron yield} The determination of neutron yield also requires the conversion efficiency of laser energy to ion energy ($\eta$) along with the convoluted IEDF as given in Eqs.(\ref{eq:neu_yield}),(\ref{eq:ave_neu_yield}) and(\ref{eq:n_ions}). In our earlier paper\cite{Hol_Neu}, we defined $\eta=N_0 E_{av}/E_{ab}$, where E\tsb{av} is average kinetic energy, E\tsb{ab} is energy absorbed by the clusters and N\tsb{0} is total number of atoms in the clusters. The total energy absorbed by the cluster is defined as the sum of total kinetic and potential energy of all particles (ions as well as electrons) along with a smaller fraction of energy required to ionize the neutral atoms of the cluster. We note that N\tsb{0} can be defined as equal to the initial number of atoms in the cluster only in the case of cluster vertical ionization (CVI)\cite{LastCVI}. In this case, the time scales of inner and outer ionization are much smaller than the time scales involved in the Coulomb explosion. At the time of Coulomb explosion, there are no inner electrons in the cluster which could have otherwise shielded the D ions undergoing the explosion. So one can safely equate the number of D ions to the number of D atoms present initially in the cluster. This kind of extension is not possible for the case when larger clusters are undergoing Coulomb explosion at lower laser intensities. For this situation, the strength of Coulomb explosion is lowered due to the presence of inner electrons in the cluster.  In our simulations, we encounter this kind of situation for large cluster size of $\langle R\rangle=200$ \AA. In particular, we have shown a significant amount of inner electron population for R\tsb{mx}=270\AA\
at I\tsb{mn}=1$\times$10\tsp{17}W/cm\tsp{2} in Fig. \ref{fig.7}d. In this kind of situation, one has to consider the effect of shielding of ion charges by inner electrons and can not equate the number of D ions to D atoms present initially in the cluster. In stead of this, the number of D ions undergoing Coulomb explosion will be equal to the total number of D ions (inner electrons + outer electrons) minus the number of inner electrons which is the outer electron population level. 

This point can be further elaborated by explicitly calculating the conversion efficiency for the two extreme cases. For the case of R\tsb{mn}=52\AA\ at I\tsb{mx}= ($\langle R_0\rangle=80$ \AA), the ratio of inner electron population to outer electron population level (N\tsb{ei}/N\tsb{eo}) is merely 1\%. Consequently, the conversion efficiencies without taking the inner electron shielding (0.621) and with electron shielding (0.615) do not differ significantly. For the case of R\tsb{mx}=270\AA\ at  I\tsb{mn}=1$\times$10\tsp{17}W/cm\tsp{2} ($\langle R\rangle=200$ \AA),  the ratio of inner electron population to outer electron population level (N\tsb{ei}/N\tsb{eo}) is merely 28.71\%. In this case, there is a significant difference between conversion efficiencies with (0.652) and without (0.840) taking into account the inner electron shielding. The other important factor is related with the calculation of final value of conversion efficiency for each average cluster size as we have considered a full range of cluster radii (R\tsb{mn}-R\tsb{mx}) and laser intensities (I\tsb{mn}-I\tsb{mx}) for each case of average sized cluster irradiated with peak spatial-temporal laser intensity of 1$\times$10\tsp{18}W/cm\tsp{2}. For one such particular case ($\langle R_0\rangle=200$ \AA\ , I\tsb{p}=1$\times$10\tsp{18}W/cm\tsp{2}), we have calculated the conversion efficiencies ($\eta$) for all cluster radii and laser intensities used for the convolution. These values of $\eta$ are shown in Table\ref{tab:table3} and finally, we take an average of all to determine the final conversion efficiency ($\eta$\tsb{RI}). 
\begin{table}
\caption{\label{tab:table3} Single cluster conversion efficiencies ($\eta$) are shown for complete cluster size (R\tsb{mn}-R\tsb{mx}) and intensity range (I\tsb{mx}--I\tsb{mn}) along with conversion efficiency averaged over cluster size ($\eta$\tsb{R}) and averaged over both cluster size and laser intensity ($\eta$\tsb{RI}). The average radius of the cluster is taken as 200\AA\ whereas the peak spatial-temporal intensity of laser is kept at 1$\times$10\tsp{18} W/cm\tsp{2}.}
\begin{ruledtabular}
\begin{tabular}{c|cccccc|c}
\backslashbox{I(W/cm\tsp{2})}{R(\AA)} & 135& 168& 190& 210& 225& 270 & $\eta$\tsb{R}\\ \hline
1$\times$10\tsp{18} W/cm\tsp{2} & 0.642 & 0.645 & 0.668 & 0.680 & 0.691 & 0.703 & 0.671\\
8$\times$10\tsp{17} W/cm\tsp{2} & 0.643 & 0.658 & 0.682 & 0.681 & 0.694 & 0.705 & 0.677\\
4$\times$10\tsp{17} W/cm\tsp{2} & 0.646 & 0.666 & 0.675 & 0.686 & 0.692 & 0.717 & 0.680\\
1$\times$10\tsp{17} W/cm\tsp{2} & 0.655 & 0.659 & 0.669 & 0.670 & 0.671 & 0.652 & 0.662\\ \hline
$\eta$\tsb{RI} & & & & & & & 0.672\\
\end{tabular}
\end{ruledtabular}
\end{table}  

\begin{figure}
\centering \includegraphics[height=3.5in,width=5.0in]{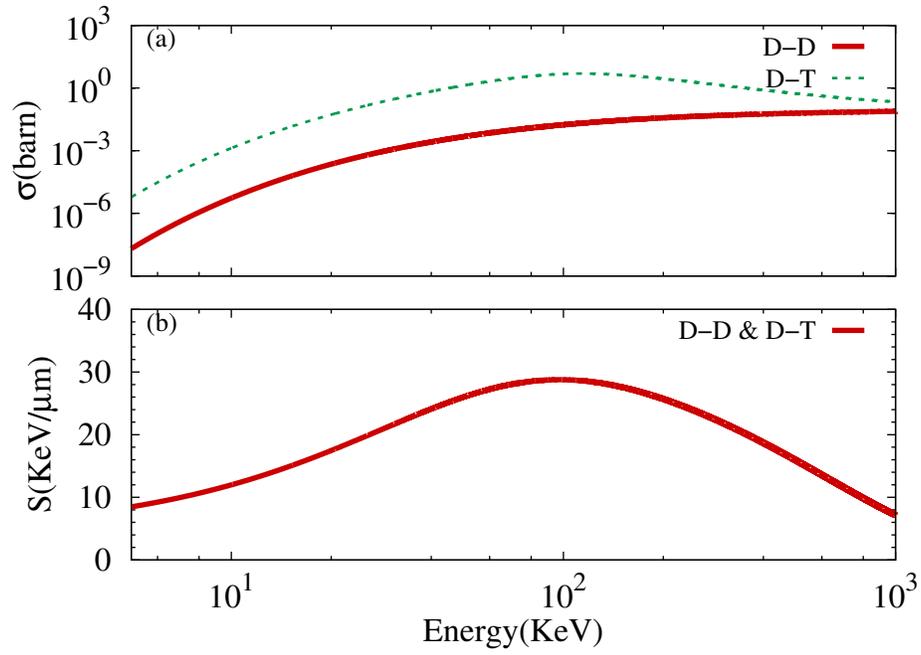}
\caption{Fusion cross section for D-D and D-T reaction (a), stopping power of D in D and D in T as a function of D ion energy for target density of $5\times 10^{22}$ cm$^{-3}$.}
\label{fig.10}
\vskip 0.5in
\hskip 4.5in G. Mishra et \emph{al.}, Fig. 10
 \end{figure}

\begin{figure}
\centering
 \includegraphics[height=3.5in,width=5.0in]{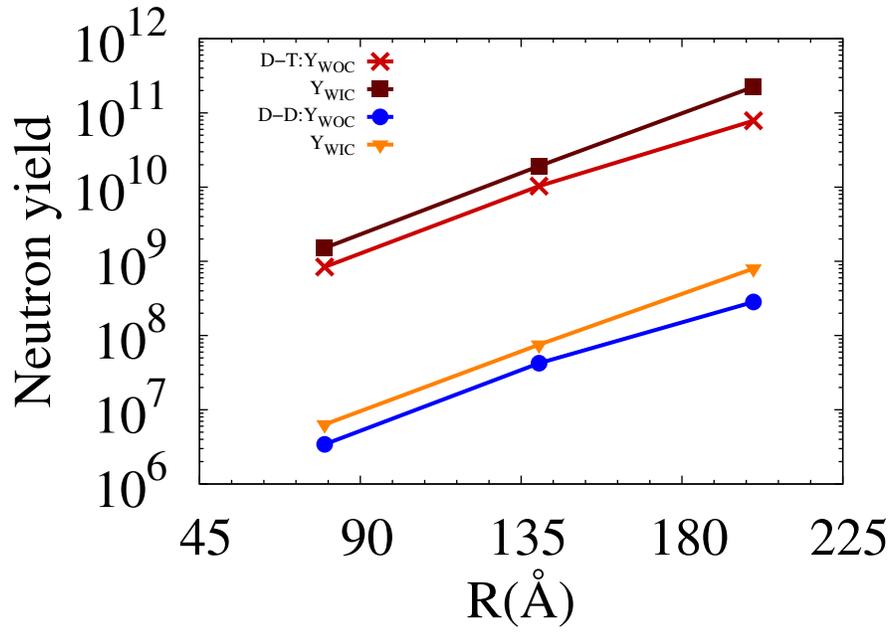}
\caption{(color on-line) Neutron yield per joule of laser energy with and without convolution for D-D and D-T. The other simulation parameters are same as used in Fig.\ref{fig.8}.}
 \label{fig.11}
\vskip 0.5in
\hskip 4.5in G. Mishra et \emph{al.}, Fig. 11
 \end{figure}

The D-D and D-T fusion cross section along with stopping power for the case of target material of density 5$\times$10\tsp{22} cm\tsp{-3} is shown in the Fig.\ref{fig.10} as a function of energy. We note from this figure that D-T fusion cross section is more than D-D where as stopping power does not change appreciably for the two cases. The resulting fusion neutron yield per joule of laser energy for these two fusion nuclear reactions is shown in Fig.\ref{fig.11} as a function of average cluster radius. It is important to note that neutron yield Y\tsb{WIC} appearing in Fig.\ref{fig.11} is calculated by using F\tsb{c} and $\eta_{RI}$ whereas neutron yield Y\tsb{WOC} in the same figure is calculated by using F and $\eta$. The yields are more for D-T than D-D due to increased fusion cross section for D-T case. When convolution is ignored, the neutron yield increases with the cluster sizes as the ion energies are increasing and the number of ions with energies close to broad maximum of fusion cross section is also increasing. When convolution is included in the determination of IEDF and $\eta$, the neutron yield approximately doubles with the case when convolution is ignored. This is due to the broadening of energy spectrum at corresponding cluster size due to the convolution effects of cluster size distribution and spatial profile of laser intensity.   

\section{Conclusion}
\label{sec:conc}
We have investigated the energetics and resulting fusion neutron yield from Coulomb explosion of Deuterium clusters (average cluster size $\langle R_0\rangle$=80,140,200\AA) driven by intense (peak spatial-temporal intensity I\tsb{p}=1$\times$ 10\tsp{18}W/cm\tsp{2}) femtosecond laser pulses. We have included the effect of cluster size and spatial laser intensity distribution in our studies by taking convolution of single cluster single intensity ion energy distribution function (IEDF) over cluster size and laser intensity. For each set of $\langle R_0\rangle$ and I\tsb{p}, a full range of cluster radii (R\tsb{mn}-R\tsb{mx}) and laser intensities (I\tsb{mx}-I\tsb{mn}) are taken for size and intensity convolution. Energy spectrum of Deuterium ions gets broadened after convolution. Size convolution significantly enhances the maximum energy and intensity convolution marginally reduces the average energy of ions compared to their single cluster single intensity counterpart. The resulting fusion neutron yield per joule of laser energy increases with cluster size and it is more for D-T than D-D fusion reaction. The double convolution also increases the neutron yield by a factor of $\sim2$ which is due to the appearance of high energy ions in the convoluted IEDF.

\section*{Acknowledgements}
The authors are thankful to Dr. N. K. Gupta for various discussions and suggestions. A.H. acknowledges the Science and Engineering Research Board, Department of Science and Technology, Government of India for funding the project SR/FTP/PS-189/2012.   

%
%

%



%

\pagebreak
 \newpage 
 
 \newpage 
 
 \newpage 
 
\newpage

 \newpage 
 
\newpage 

 \newpage 
 
 \newpage 
 
 \newpage 
 
 \newpage 
 
 \newpage  
\end{document}